\documentclass[a4paper,12pt]{article}
\usepackage{graphicx,epsfig,cite}
\usepackage{amsfonts}
\usepackage{amsmath}
\usepackage{amssymb}
\usepackage{longtable}
\usepackage{setspace}
\usepackage[usenames]{color}
\usepackage{colortbl}

\newlength{\abc}
\title{\bf The Features Radiation of a Charged Particle Moving Along
an Arc of a Circle}
\author{S.E.~Boychenko, V.B.~Tlyachev\footnotemark[1]\\[2mm]
\footnotemark[1]~Adygea State University, 385000, Maykop,
Pervomayskaya str. 208, Russia, \\tlyachev@adygnet.ru}
\date{}

\begin{document}
\maketitle
\begin{center}
\textbf{Abstract}

\small{A detailed analysis is given of the angular distribution of
an charged particle moving along the arc of a circle.
The areas of different angular distribution behavior are highlighted and studied.}

\end{center}

Recently discovered new properties of synchrotron radiation (SR)
(see [1] and subsequent works [2-5]) gave a new impulse to the
development of the SR theory. One of the questions of the SR
theory is the study of the radiation characteristics of an
electron moving along an arc of a circle. Such motion is
considered as an integral part of more complex motion
trajectories, available, for example, in synchrotrons, undulators
or wigglers.

The main theoretical results in this area were obtained by using
classical methods of electrodynamics in the works of Bagrov,
Ternov, Fedosov [6-7].

A numerical analysis of the radiation structure was carried out
and asymptotic formulas were in the articles [8-10] and in
monograph [11]. The analysis was carried out for the particle
velocity $v=\beta c,$ where $c$ is the speed of light and
$\beta>\sqrt{2/5}.$

In 1980s the results obtained for this speed range were considered
sufficient (the a non relativistic and ultra relativistic cases) .
At the same time, a full study required significant computing
power. Modern computer tools have solved this problem.

In this paper, a detailed analysis of the angular distribution of
radiation was provided for charged particles moving in a plane
with a constant velocity $v=\beta c.$ Particle moves under the arc
of a circle of opening angle $2\gamma$ in the plane of the orbit
(Fig.~1.). The beta value takes values from $0$ to $1.$

\begin{figure}[htbp!]
\begin{center}
\epsfig{file=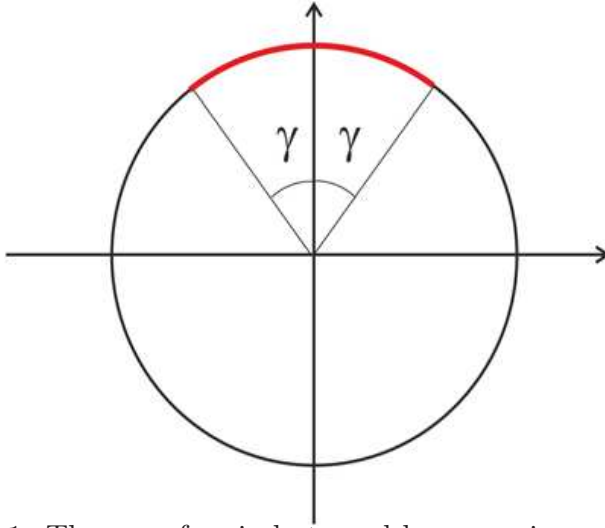,width=8cm,height=7cm}
\renewcommand{\figurename}{Fig.}\vspace{-0.6cm}
\caption{The arc of a circle traced by a moving
particle}
\label{fig5.1}
\end{center}
\end{figure}

We used the expressions for the angular distribution of radiation
obtained in [6,7] by the classical electrodynamics methods. For
numerical analysis we use exact expressions distribution of the
average power of radiation integrated over the frequency for the
$\sigma$-- and $\pi$-- components of linear polarization in the
form:
\begin{equation}
    \label{eq4.1:ref}
    \operatorname d\varepsilon=W_0 T F(\beta,\gamma;\Theta,\varphi)\operatorname d\Omega,\;F=F_\sigma+F_\pi,
\end{equation}
\begin{equation}
    \label{eq4.2:ref}
   F_{\sigma,\pi}(\beta,\gamma;\Theta,\varphi)=\Phi_{\sigma,\pi}^0+\Phi_{\sigma,\pi}^0(\varphi+\gamma)+\Phi_{\sigma,\pi}^1(\varphi-\gamma),
\end{equation}
where $W_0$ --- is total power synchrotron radiation of an
electron moving in a circle of radius $R,$  $T$ --- is time of
movement on an arc of circle, $\Theta,\varphi$ --- are the polar
and azimuthal angles of spherical coordinates, the direction of
observation of radiation characterizing.

The functions $\Phi_{\sigma, \pi}^0(\beta,\Theta)$ describe the
angular distribution of the average per revolution of synchrotron
radiation:
\begin{eqnarray}
    \label{eq4.3:ref}
   \Phi_\sigma^0(\beta,\Theta)=\frac{3\left(1-\beta^2\right)^2\left(4+3\mu^2\right)}{64\pi\left(1-\mu^2\right)^{5/2}},\;
   \Phi_\pi^0(\beta,\Theta)=\frac{3\left(1-\beta^2\right)^2\left(4+\mu^2\right)\cos^2\Theta}{64\pi\left(1-\mu^2\right)^{7/2}},\;
\end{eqnarray}
where $\mu=\beta\sin\Theta,\;0\leq\beta\leq 1.$
\begin{eqnarray}\label{eq4.4:ref}
\Phi_\sigma^1(x)=Q(1-\mu^2)\Bigg\{\frac{6\mu
p^4(4+3\mu^2)}{\left(1-\mu^2\right)^{1/2}}arctg\left(\frac{\mu\sin
x}{p+\left(1-\mu^2\right)^{1/2}}\right)+ \\ \nonumber
+\left[\left(23\mu^2-2\right)p^3 +\left(9\mu^2-2\right)\left(1-\mu^2\right)p^2-2\left(1-\mu^2\right)^2p+6\left(1-\mu^2\right)^3\right]\sin x\Bigg\}, \\
\Phi_\pi^1(x)=Q\cos^2\Theta\Bigg\{\frac{6\mu
p^4(4+\mu^2)}{\left(1-\mu^2\right)^{1/2}}arctg\left(\frac{\mu\sin
x}{p+\left(1-\mu^2\right)^{1/2}}\right)+ \\ \nonumber
+\left[\left(2+13\mu^2\right)p^3+\left(3\mu^2+2\right)\left(1-\mu^2\right)p^2+2\left(1-\mu^2\right)^2p-6\left(1-\mu^2\right)^3\right]\sin
x\Bigg\}, \nonumber
\end{eqnarray}
where
$Q=\left(1-\beta^2\right)\left[128\pi\gamma\mu\left(1-\mu^2\right)^3p^4\right]^{-1},\;p=1-\mu\cos
x.$

The formulas (1)-(5) describe a continuous transition from the
instantaneous angular distribution of synchrotron radiation to the
average per revolution when the opening angle $2\gamma$ changes
from $0$ to $\pi.$

Since the angular distribution is symmetric with respect to the
plane $y=0,$ in numerical analysis we restrict ourselves to
considering angles $0\leq\varphi\leq\pi.$

A numerical analysis of the $\sigma$--component linear
polarization of the radiation shows that with increasing particle
velocity the radiation concentration occurs in the orbit plane.
This is completely consistent with the known results from the SR
theory. Therefore further we consider radiation in the orbit plane
$\Theta=\pi/2.$ In this case, the angular distribution is
conveniently represented as a function
\begin{equation}
    \label{eq4.5:ref}
  \chi(\varphi)=F_\sigma\left(\beta,\gamma,\frac\pi2,\varphi\right),\;0\leq\gamma\leq\pi.
\end{equation}

The function $\chi(\varphi)$ always has extrema at the points
$\varphi=0,\varphi=\pi$ for any $\beta, \gamma.$ Besides this
function may have extrema at points
$\varphi=\varphi_k,\;(k=1,2,3,4).$

The functions $\varphi_k$ satisfy inequalities
\begin{equation}
    \label{eq4.6:ref}
  0\leq\varphi_1<\varphi_2<\gamma<\varphi_3<\varphi_4<\pi
\end{equation}

Let's consider these functions in more detail. The functions
$\varphi_k$ can be obtained as solutions of the equation
$\chi'(\varphi)=0.$ The solutions of this equation substantially
depend on the particle velocity $\beta.$ The threshold value here
is $\beta=2/5.$ When passing through this value, significant
changes are observed in the structure of the angular distribution
associated with the appearance of two additional extrema. For
$0<\beta\leq 2/5$ there are only two internal extrema
$\varphi_2(\beta,\gamma)$ and $\varphi_3(\beta,\gamma).$ For
$2/5<\beta<1$ the radical expression in formula for
$\chi'(\varphi)$ becomes positive and extrema appear
$\varphi_1(\beta,\gamma)$ and $\varphi_4(\beta,\gamma).$ Thus we
consider two cases: $0<\beta\leq 2/5$ and $2/5<\beta<1.$

\begin{center}
\textbf{Angular distribution at $0<\beta\leq 2/5$}
\end{center}

Solving the equation $\chi'(\varphi)=0$ for
$0<\beta\leq\frac{2}{5}$ and, discarding the roots $\varphi=0,
\varphi=\pi,$ we obtain two solutions $\varphi_2(\beta,\gamma)$
and $\varphi_3(\beta,\gamma).$ These functions are considered as
implicit functions for the fixed values of  $\beta.$ See the
Fig.~2.
\begin{figure}[tbh!]
\begin{center}
\includegraphics [scale=0.65]{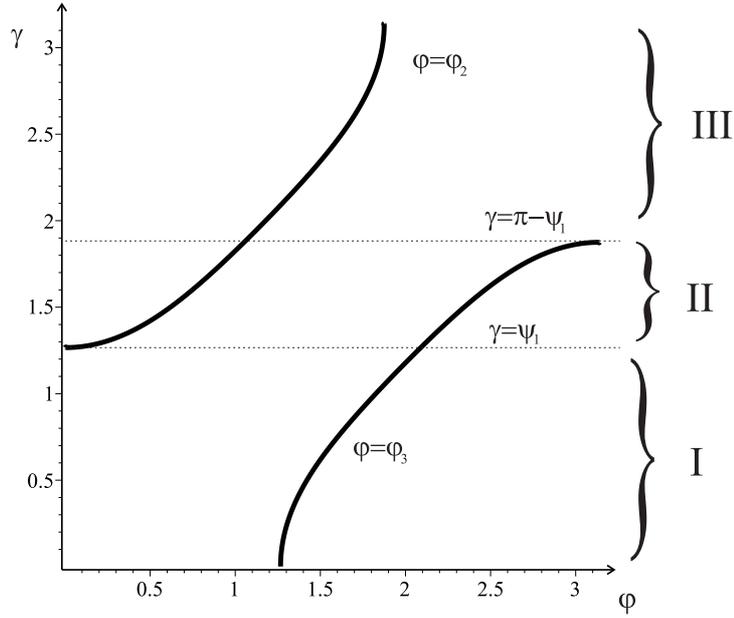}
\renewcommand{\figurename}{Fig.}
\vspace{-0.6cm}\caption{The structure of extrema $\varphi_2,$
$\varphi_3$ of the function $\chi(\varphi)$ for $\beta<2/5$}
\label{Fig5.2}
\end{center}
\end{figure}

Let us consider the function $\psi_1(\beta)$ at which
$\varphi_2(\beta,\gamma)$ reaches a minimum. This function is
exactly found and it is equal to $\psi_1(\beta)=arccos(\beta)).$
In this case, we have three areas at which the behavior of the
function $\chi(\varphi)$ is uniquely determined. The distribution
of extremes is shown on the Fig.~2. At low values of beta (less
than $2/5$) there are only three qualitatively different extremum
locations (Fig. 3-5).
\begin{figure}[htbp]
\epsfig{file=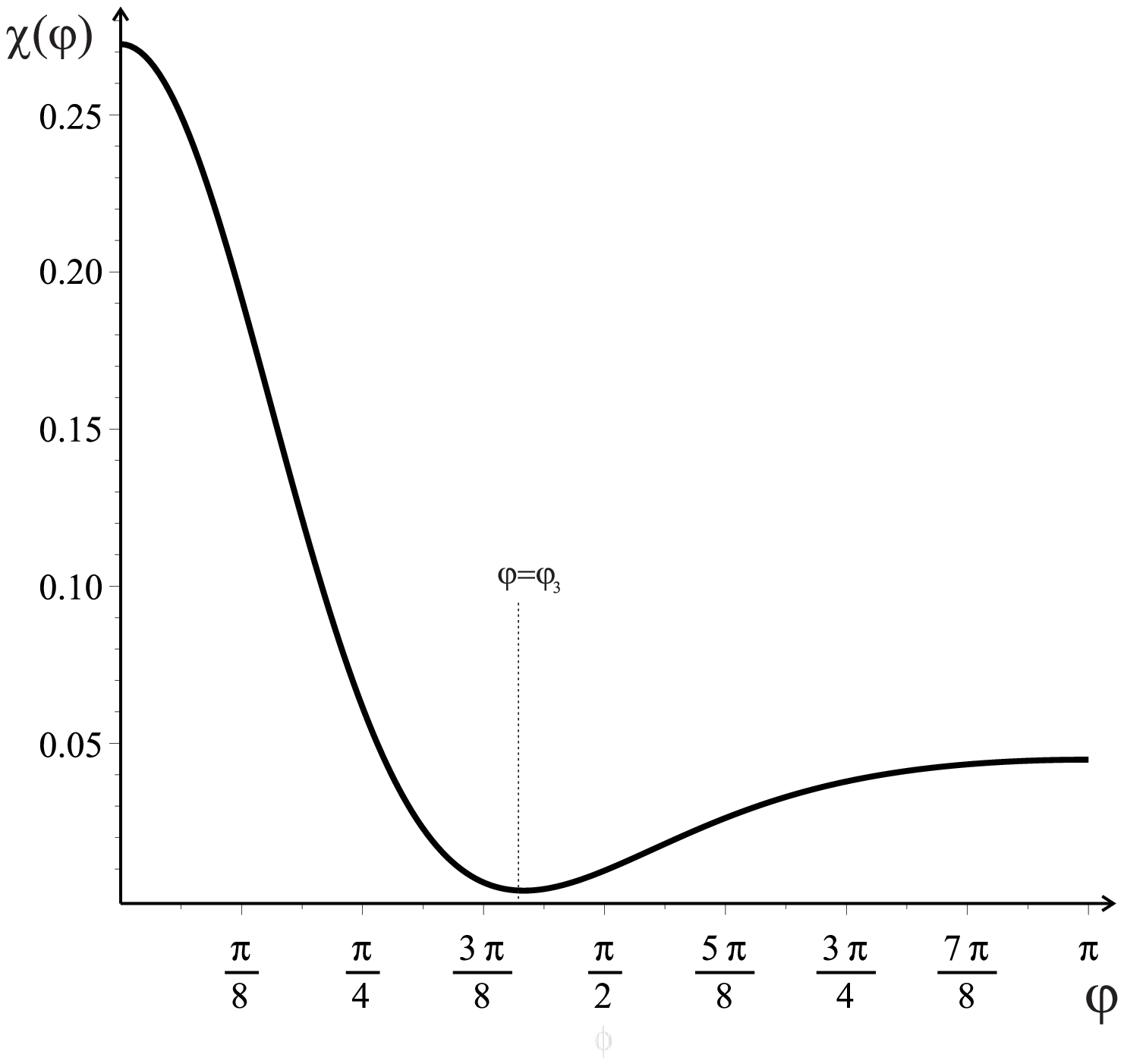,width=8cm,height=8cm} \hfill
\epsfig{file=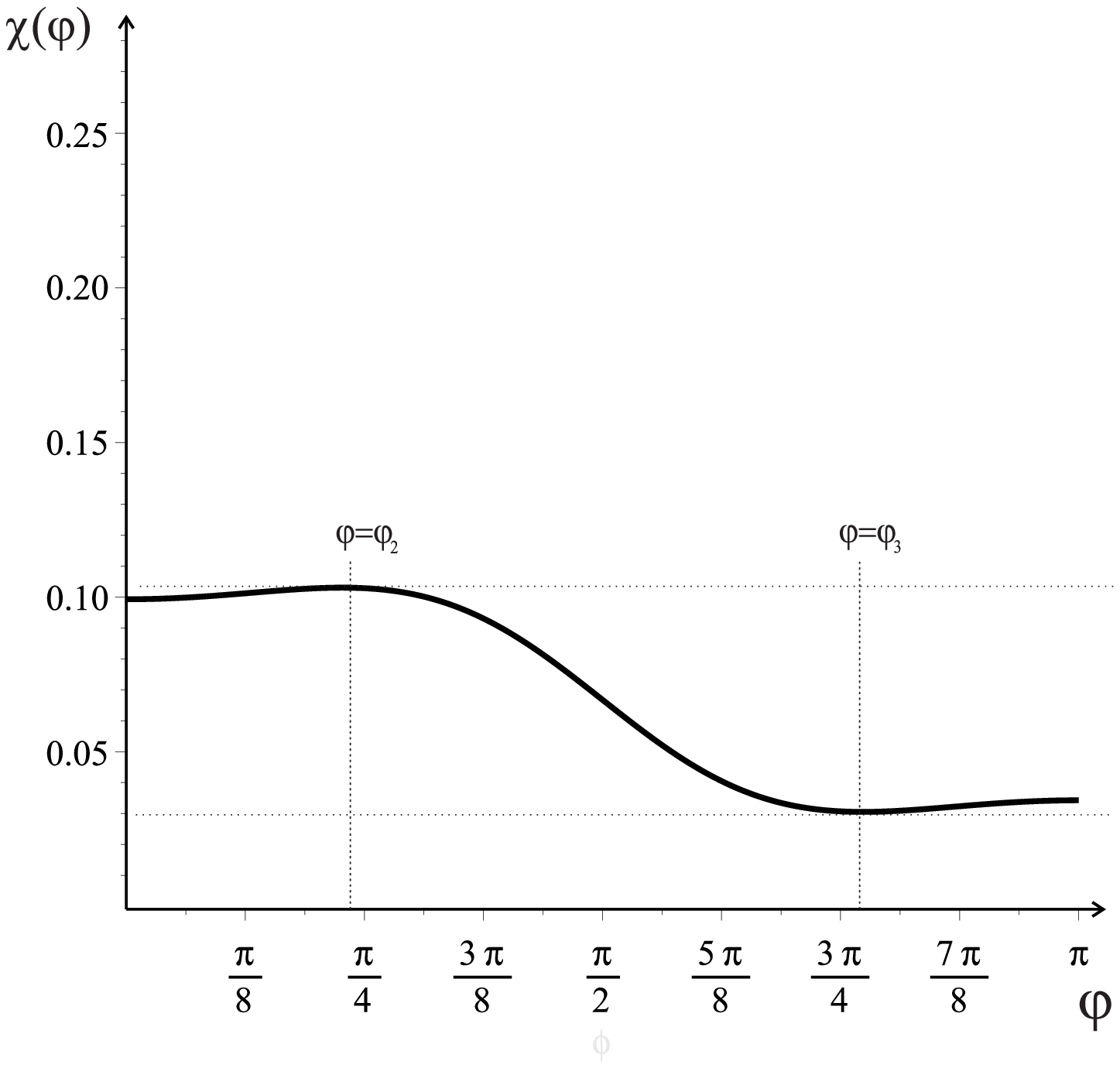,width=8cm,height=8cm} \hfill
%\begin{center}
\renewcommand{\figurename}{Fig.}
\parbox[t]{0.41\textwidth}{
\caption{$\chi(\varphi)$ for zone I $\beta=0.3,$ $\gamma=0.26,$
$\psi_1=1.27,$ $\varphi_3$ --- global minimum}}\label{fig5.3.1}
%\end{center}
\hfill
%\begin{center}
\parbox[t]{0.41\textwidth}{\caption{$\chi(\varphi)$ for zone II $\beta=0.3,$
$\gamma=\pi/2,$ $\psi_1=1.27,$ $\varphi_2$ --- global maximum,
$\varphi_3$ --- global minimum}} \label{fig5.3.2}
%\end{center}
\end{figure}

\begin{figure}[tbh!]
\begin{center}
\includegraphics [scale=0.6]{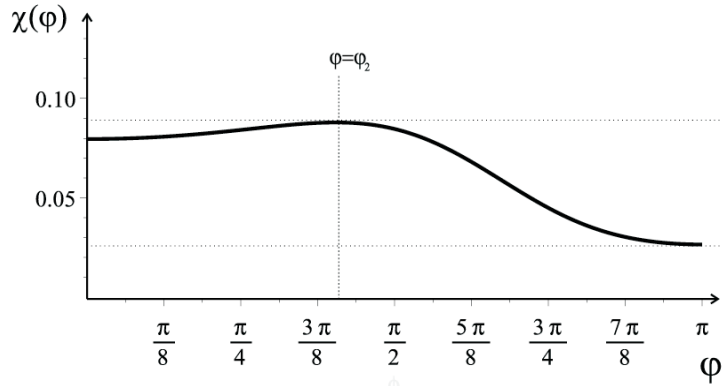}
\renewcommand{\figurename}{Fig.}
\vspace{-0.6cm}\caption{$\chi(\varphi)$ for zone III $\beta=0.3,$
$\gamma=2.09,$
$\psi_1=1.27,$ $\varphi_2$ --- global maximum} \label{fig5.3.3}%Fig.5
\end{center}
\end{figure}

\newpage

\begin{center}
\textbf{Angular distribution at  $2/5<\beta<1$}
\end{center}

The structure of the angular distribution becomes more complex
with increasing particle velocity. The new additional extrema
$\varphi_1$ and $\varphi_4$ appear. This is due to the fact that
the equation $\chi'(\varphi)=0$ has new roots that does not exist
for $\beta<2/5.$ We define the auxiliary function $\psi_2(\beta)$
as the maximum of $\psi_4.$ This maximum value is reached at
$\varphi=\pi.$ Thus, we obtain the exact value
\begin{equation}
    \label{eq4.8:ref}
  \psi_2(\beta)=arccos\frac{5\beta^2-2}{3\beta}.
\end{equation}

It should be noted that a study of the radiation of a charge
moving along an arc of a circle in the relativistic case was
carried out in [8,9]. However, in [8,9] the threshold value
$\beta=2/5$ was not found and the results were obtained under the
assumption that $\beta$ is greater than $\sqrt{2/5}.$ For the
remaining values of $\beta$ the angular distribution was not
studied.

Numerical calculations made it possible to determine five areas
related to the angle $\gamma,$ where the radiation structure
changes significantly with increasing $\beta.$

The key intervals here are: $2/5<\beta<1/2,$
$1/2<\beta<\sqrt{2/5}$ and $\beta>\sqrt{2/5}.$ The angular
structure of radiation for $2/5<\beta<1/2$ is shown in Fig.~6.
\begin{figure}[tbh] %Fig.6
\begin{center}
\includegraphics [scale=0.65]{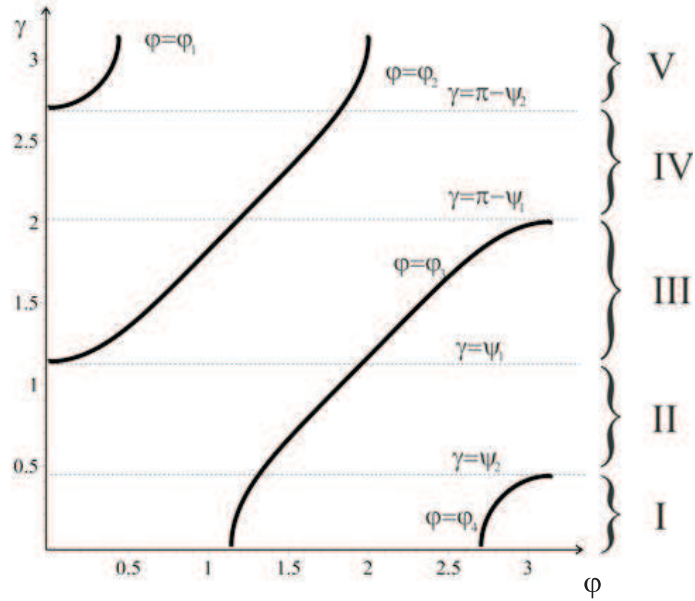}
\renewcommand{\figurename}{Fig.}
\vspace{-1.2cm}\caption{The structure of extrema $\varphi_k$ of
the function $\chi(\varphi)$ for $2/5<\beta<1/2.$} \label{fig5.4}
\end{center}
\end{figure}

In the range $2/5<\beta<1/2$ the condition
$\psi_1(\beta)>\psi_2(\beta)$ is satisfied. When the particle
reaches the velocity $\beta=1/2,$ the equality holds
$\psi_1(\beta)=\psi_2(\beta).$ With a further increase of the
particle velocity, it turns into an inequality
$\psi_1(\beta)<\psi_2(\beta).$  See Fig. 7. This fact explains the
changes in angular distribution structure.

\begin{figure}[htbp!]
\epsfig{file=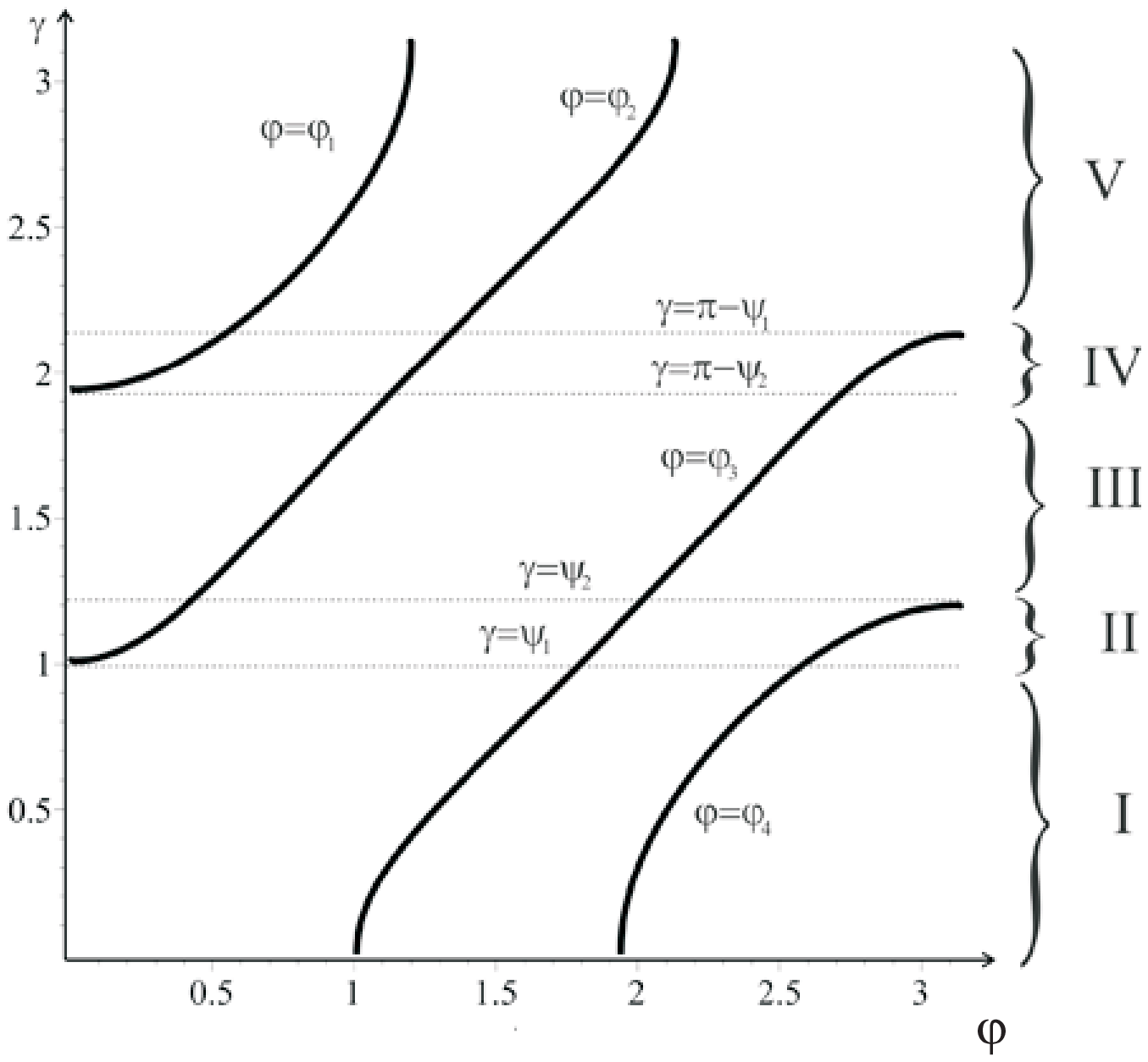,width=8cm,height=8cm} \hfill
\epsfig{file=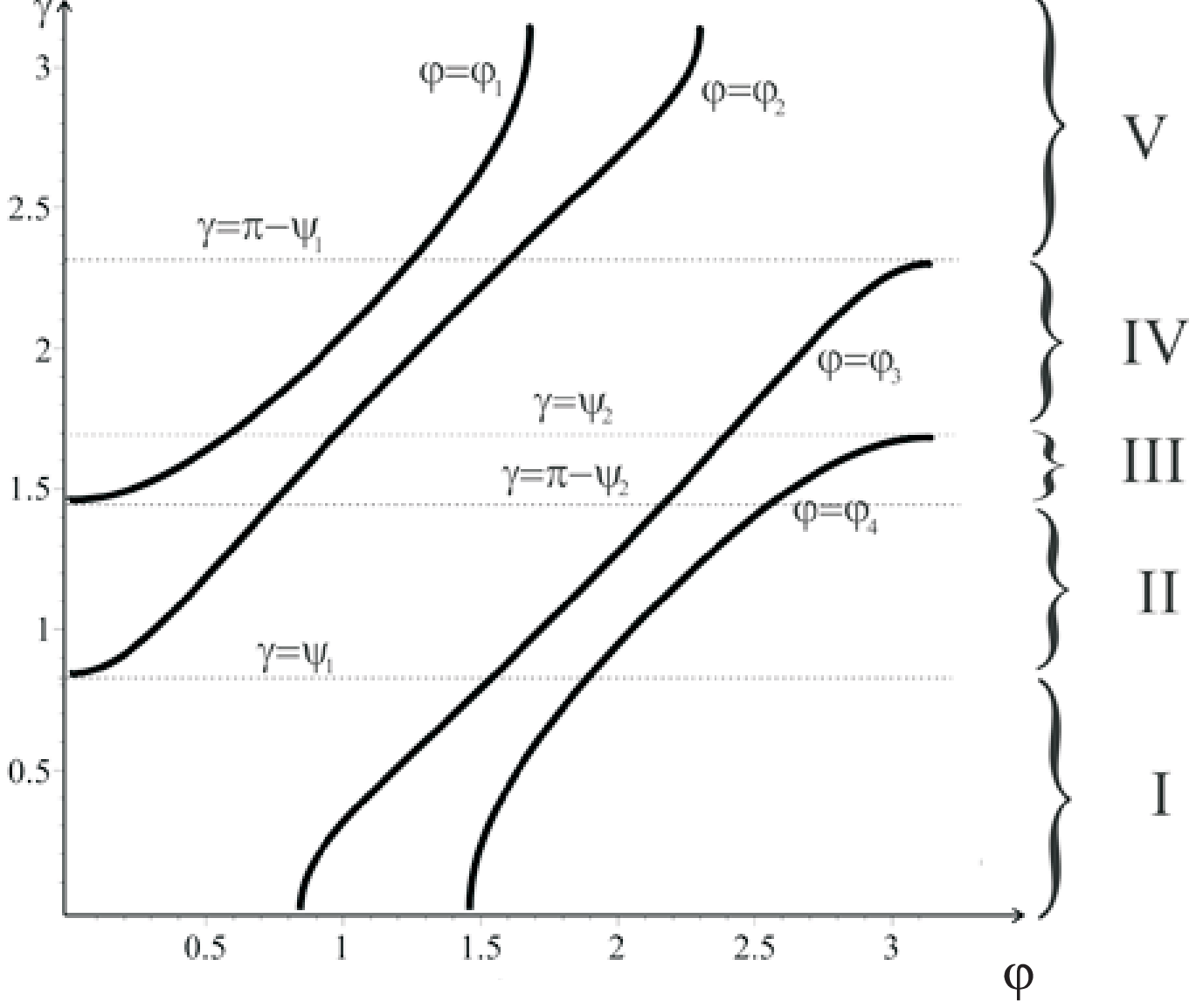,width=8cm,height=8cm} \hfill
%\begin{center}
\renewcommand{\figurename}{Fig.}\vspace{-1.6cm}
\parbox[t]{0.41\textwidth}{\caption{Structure of extrema $\varphi_k$ of the function
$\chi(\varphi)$ for $1/2<\beta<\sqrt{2/5}$}}\label{fig5.5}
%\end{center}
\hfill
%\begin{center}
\parbox[t]{0.41\textwidth}{\caption{The structure of extrema $\varphi_k$
of the function $\chi(\varphi)$ for $\sqrt{2/5}<\beta<1$}}
\label{fig5.6}
%\end{center}
\end{figure}

The next significant change in the structure of radiation occurs
when the velocity of particle $\beta>\sqrt{2/5}.$ In this case
zone (area) IV is located above zone III in contrast to Fig.~6,
where the picture is opposite. See Fig.~8.

The behavior of the function $\chi(\varphi)$ in this
 case ($\beta>\sqrt{2/5}$) was studied in [9]. The numerical
 analysis shows that for $0\leq\gamma<\psi_1$ (zone I in Fig.~8)
 the point $\varphi=0$ is the global (absolute) maximum, the
point $\varphi=\varphi_3$ --- the global (absolute) minimum, the
point $\varphi=\varphi_4$ --- the local maximum and the point
$\varphi=\pi$ is the local minimum.

For $\psi_1\leq\gamma<\pi-\psi_2$ (zone II in Fig.~8) the point
$\varphi=0$ is the local minimum, the point $\varphi=\varphi_2$ is
the global (absolute) maximum, the point $\varphi=\varphi_3$
--- the global (absolute) minimum, the point $\varphi=\varphi_4$ ---
the local maximum and the point $\varphi=\pi$ is the local
minimum.

For $\pi-\psi_2\leq\gamma<\psi_2$ (zone III in Fig.~8) the
function $\chi(\varphi)$ has in the points $\varphi=\varphi_1,
\varphi_3, \pi$ the minimums and in the points $\varphi=0,
\varphi_2, \varphi_4$ --- maximums.

For $\psi_2\leq\gamma<\pi-\psi_1$ (zone IV in Fig.~8) the function
$\chi(\varphi)$ has the five extreme points. The point $\varphi=0$
is the global (absolute) maximum, the point $\varphi=\varphi_1$ is
the local minimum, the point $\varphi=\varphi_2$ --- the local
maximum, the point $\varphi=\varphi_3$ --- global (absolute)
minimum and the point $\varphi=\pi$ is the local maximum.

For $\pi-\psi_1\leq\gamma<\pi$ (zone V in Fig.~8) the point
$\varphi=0$ is the global (absolute) maximum, the point
$\varphi=\varphi_1$ is the local minimum, the point
$\varphi=\varphi_2$ --- the local maximum and the point
$\varphi=\pi$ is the global (absolute) minimum.

For velocities $\beta$ close to unity, asymptotic formulas for
$\varphi_k$ were obtained in [9]
$$
\varphi_{1,2}(\beta,\gamma)=\gamma+\sqrt{\varepsilon_0}\pm\varepsilon_0^2
(2\sqrt{2}\sin^3\gamma)^{-1}, \eqno(10)
$$
$$
\varphi_{3,4}(\beta,\gamma)=\gamma-\sqrt{\varepsilon_0}\pm\varepsilon_0^2
(2\sqrt{2}\sin^3\gamma)^{-1}, \eqno(11)
$$
where $\varepsilon_0 =1-\beta^2.$

The numerical analysis made it possible to clarify the action
areas of asymptotic formulas (10)--(11). We can conclude that
asymptotic expressions have a large error for small values of the
angle $\gamma$ and for $\gamma$ near to $\pi.$ The formulas (10)
and (11) give good results in the range of angles
$\pi/4<\gamma<3\pi/4$ for $\beta > 0.9.$  This can be seen in more
detail from the tables 1-4 (Appendix). The error of
$\delta_{\varphi_i}$ of asymptotic formulas was calculated using
the formula
\begin{equation}
    \label{eq4. 12: ref}
  \delta_{\varphi_i}=\frac{\left|\varphi_{real}-\varphi_{asympt}\right|}{\varphi_{real}},\;i=1\ldots 4,
\end{equation}
where $\varphi_{real}$ is calculated as the solution of the
equation $\chi'(\varphi)=0$. The results are presented in table 5.

To summarize we can say that numerical analysis and computer
simulations allowed us to investigate in sufficient detail the
structure of the angular distribution of particle radiation at all
kinds of velocity values $0<\beta<1.$

\begin{center}
\textbf{References}
\end{center}

\noindent [1] Bagrov V.G. Special features of the angular
distribution of synchrotron radiation power of discrete spectral
harmonics // Russian Physics Journal. 2008. 51:335. \vskip 0.35cm

\noindent [2] Bagrov V.G., Gitman D.M. et al. Effective spectrum
width of the synchrotron radiation. Eur. Phys. J. C 75(11), 555
(2015). \vskip 0.35cm

\noindent [3] Bagrov V.G., Gitman D.M. et al. Dependence of
effective spectrum width of synchrotron radiation on particle
energy // Eur. Phys. J. C (2017) 77: 345.\vskip 0.35cm

\noindent [4] Bagrov V.G. On the Wave Zone of Synchrotron
Radiation // arXiv:1806.08491 [physics.class-ph] (2018)\vskip
0.2cm

\noindent [5] Loginov A.S., Saprykin A.D. Numerical Analysis of
the Location of Spectral Maxima of Synchrotron Radiation
Polarization Components in Classical Theory // Russian Physics
Journal 2018. Vol. 61. No 3. P. 534-539. \vskip 0.35cm

\noindent [6] Bagrov V.G., Ternov I.M., Fedosov N.I. Radiation of
relativistic electrons moving along the arc of a circle // JETP.
1982. Vol. 55, No. 5. P. 835-838.\vskip 0.2cm

\noindent [7] Bagrov V.G., Fedosov N.I., Ternov I.M. Radiation of
relativistic electrons moving in an arc of a circle // Phys. Rev.
1983. Vol. 28. No 10. P. 2464-2472.\vskip 0.35cm

\noindent [8] Bagrov V.G., Razina G.K.at al. Singularities of
angular spectral characteristics of the radiation of an electron
moving along the ARC of a circle // Soviet Physics Journal. 1982.
Vol. 25. No 7. P. 648-651.\vskip 0.35cm

\noindent [9] Bagrov V.G., Razina G.K.at al. Numerical analysis of
spectral distribution of power radiation of an electron moving
along the ARC of a circle. 30 p. VINITI 30.07.85, No 5599-85.
Russian Physics Journal (1985). \vskip 0.35cm

\noindent [10] Kargin Y.N., Razina G.K., Fedosov N.I. Spectral
characteristics of radiation of a nonrelativistic electron moving
along the arc of a circle // Russian Physics Journal. 1985. Vol.
28. No 4.  P. 305-308.  \vskip 0.35cm

\noindent [11] Bagrov V.G., Bisnovaty-Kogan G.S., Bordovitsyn V.A.
et al. Theory of radiation of relativistic particles. Ed. V.A.
Bordovitsyna. Moscow: Fizmatlit, 2002. 575 p. (In Russian)

\newpage

\begin{center}
\textbf{Appendix}
\end{center}

\begin{small}
\begin{spacing}{0.05}
\begin{flushright}
%\caption{Table 1}\\
\textit{Table 1}
\end{flushright}
\begin{center}
%\caption{Table 1}\\
{$\varphi_{1}(\beta, \gamma)$}
\end{center}
\end{spacing}

\begin{table}[h!]
\begin{tabular}{|l|l|l|l|l|l|l|l|l|}
\hline
   \begin{tabular}{ll}
 & $\gamma$\\
 $\beta$ &
\end{tabular}  & & $\dfrac{\pi}{16}$ & $\dfrac{\pi}{8}$ & $\dfrac{\pi}{4}$ & $\dfrac{\pi}{2}$ & $\dfrac{3\pi}{4}$ & $\dfrac{7\pi}{8}$ & $\dfrac{15\pi}{16}$ \\ \hline
{0.7}{}{} & $\varphi_{real}$ & 1.386
& 1.474  & 1.763 & 2.5979 & undef.  & undef.   & undef.  \\
\cline{2-9}
                  & $\varphi_{asympt}$ & 13.295 & 2.748  &  1.760  & 2.377 & 3.3304  & 5.1039 & 16.044 \\ \hline
{0.8}{}{} & $\varphi_{real}$ & 1.087 & 1.192 & 1.510 & 2.2903 &
undef. & undef.  & undef. \\ \cline{2-9}
                  & $\varphi_{asympt}$ & 6.967 & 1.810 & 1.515 & 2.2166 & 3.0857 & 4.166 & 9.7162 \\ \hline
{0.9}{}{} & $\varphi_{real}$ & 0.765 & 0.902 &
\cellcolor{Gray} 1.259 & \cellcolor{Gray} 2.0376 & \cellcolor{Gray} 2.910 & undef. & undef. \\
\cline{2-9}
                  & $\varphi_{asympt}$ & 2.351 & 1.056 & \cellcolor{Gray} 1.257 & \cellcolor{Gray} 2.0194 & \cellcolor{Gray} 2.828 & 3.4125 & 5.1000 \\ \hline
{0.95}{}{} & $\varphi_{real}$ & 0.569 & 0.731 &
\cellcolor{Gray} 1.109 & \cellcolor{Gray} 1.8921 & \cellcolor{Gray} 2.6900 & undef. & undef. \\
\cline{2-9}
                  & $\varphi_{asympt}$ & 0.961 & 0.765 & \cellcolor{Gray} 1.107 & \cellcolor{Gray} 1.8864 & \cellcolor{Gray} 2.678 & 3.1211 & 3.7101 \\ \hline
{0.99}{}{} & $\varphi_{real}$ & 0.3444 & 0.5357 &
\cellcolor{Gray} 0.9273 & \cellcolor{Gray} 1.71219 & \cellcolor{Gray} 2.49826 & undef. & undef. \\
\cline{2-9}
                  & $\varphi_{asympt}$ & 0.356 & 0.536 & \cellcolor{Gray} 0.9269 & \cellcolor{Gray} 1.7120 & \cellcolor{Gray} 2.49766 & 2.8925 & 3.1052 \\ \hline
{0.999}{}{} & $\varphi_{real}$ & 0.24121 & 0.4375 &
\cellcolor{Gray} 0.83013 & \cellcolor{Gray} 1.61553 & \cellcolor{Gray} 2.40101 & 2.7937 & undef. \\
\cline{2-9}
                  & $\varphi_{asympt}$ & 0.24125 & 0.4374 & \cellcolor{Gray} 0.83011 & \cellcolor{Gray} 1.61551 & \cellcolor{Gray} 2.40091 & 2.7936 & 2.9901 \\ \hline
\end{tabular}
\end{table}
\end{small}

\begin{small}
\begin{spacing}{0.05}
\begin{flushright}
\textit{Table 2}
\end{flushright}
\begin{center}
%\caption{Table 2}\\
{$\varphi_{2}(\beta, \gamma)$}
\end{center}
\end{spacing}

\begin{table}[h!]
\begin{tabular}{|l|l|l|l|l|l|l|l|l|}
\hline
\begin{tabular}{ll}
 & $\gamma$\\
 $\beta$ &
\end{tabular} &  & $\pi/16$ & $\pi/8$ & $\pi/4$ & $\pi/2$ & $3\pi/4$ & $7\pi/8$ & $15\pi/16$ \\ \hline
{0.7}{}{} & $\varphi_{real}$ & 0.8675 & 1.0447 & 1.4689 & 2.2601 &
undef. & undef.  & undef. \\ \cline{2-9}
                  & $\varphi_{asympt}$ & -11.4743 & -0.5340 & 1.2394 & 2.1930 & 2.8102 & 1.8222 & -8.7254 \\ \hline
{0.8}{}{} & $\varphi_{real}$ & 0.7375 & 1.1924 & 1.3689 & 2.1627 &
2.8939 & undef. & undef. \\ \cline{2-9}
                  & $\varphi_{asympt}$ & -5.3746 & 0.1751 & 1.2558 & 2.12498 & 2.8266 & 2.5313 & -2.6257 \\ \hline
{0.9}{}{} &$\varphi_{real}$  & 0.5804 & 0.7998 & 1.2167 & 2.0080 &
2.7618 & undef. & undef. \\ \cline{2-9}
                  & $\varphi_{asympt}$ & -1.0867 & 0.6008 & 1.1852 & 1.9939 & 2.7560 & 2.9570 & 1.6622 \\ \hline
{0.95}{}{} & $\varphi_{real}$ & 0.4759 & 0.6924 & 1.0969 & 1.8848
& 2.6606 & 3.0204 & undef. \\ \cline{2-9}
                  & $\varphi_{asympt}$ & 0.0560 & 0.6450 & 1.0881 & 1.8797 & 2.6589 & 3.0012 & 2.8049 \\ \hline
{0.99}{}{} & $\varphi_{real}$ & 0.3321 & 0.5328 & 0.9266 & 1.7125
& 2.4973 & 2.8945 & undef. \\ \cline{2-9}
                  & $\varphi_{asympt}$ & 0.3186 & 0.5313 & 0.9261 & 1.7117 & 2.4969  & 2.8875 & 3.0675 \\ \hline
{0.999}{}{} & $\varphi_{real}$ & 0.2412 & 0.4375 & 0.83013 &
1.61551 & 2.4009 & 2.7937 & undef. \\ \cline{2-9}
                  & $\varphi_{asympt}$ & 0.2409 & 0.4374 & 0.83010 & 1.61551 & 2.4009 & 2.7936 & 2.98976 \\ \hline
\end{tabular}
\end{table}
\end{small}

\newpage

\begin{small}
\begin{spacing}{0.05}
\begin{flushright}
%\caption{Table 3}\\
\textit{Table 3}
\end{flushright}
\begin{center}
%\caption{Table 3}\\
{$\varphi_{3}(\beta, \gamma)$}
\end{center}
\end{spacing}
\begin{table}[h!]
\begin{tabular}{|l|l|l|l|l|l|l|l|l|}
\hline
 \begin{tabular}{ll}
 & $\gamma$\\
 $\beta$ &
\end{tabular} &  & $\pi/16$ & $\pi/8$ & $\pi/4$ & $\pi/2$ & $3\pi/4$ & $7\pi/8$ & $15\pi/16$ \\ \hline
{0.7}{}{} & $\varphi_{real}$ & undef. & undef. & 0 & 0.8815 &
1.6727 & 2.0969 & 2.2741  \\ \cline{2-9}
                  & $\varphi_{asympt}$ & 11.867 & 1.3194 & 0.3314 & 0.9486 & 1.9022 & 3.6756 & 14.6159 \\ \hline
{0.8}{}{} & $\varphi_{real}$ & undef. & undef. & 0.2477 & 0.9789 &
1.7727 & 2.2014 & 2.2903  \\ \cline{2-9}
                  & $\varphi_{asympt}$ & 5.7673 & 0.6103 & 0.3150 & 1.0166 & 1.8858 & 2.9665 & 8.5162  \\ \hline
{0.9}{}{} &$\varphi_{real}$  & undef. & undef. & 0.3798 & 1.1336 &
1.9249 & 2.3417 & 2.5612 \\ \cline{2-9}
                  & $\varphi_{asympt}$ & 1.4794 & 0.1846 & 0.3856 & 1.1477 & 1.9564 & 2.5407 & 4.2283 \\ \hline
{0.95}{}{} & $\varphi_{real}$ & undef. & 0.1212 & 0.4810 & 1.2568
& 2.0447 & 2.4492 & 2.6657 \\ \cline{2-9}
                  & $\varphi_{asympt}$ & 0.3367 & 0.1404 & 0.4827 & 1.2619 & 2.0535 & 2.4966 & 3.0856 \\ \hline
{0.99}{}{} & $\varphi_{real}$ & 0 & 0.2547 & 0.6443 & 1.4294 &
2.2150 & 2.6088 & 2.8095 \\ \cline{2-9}
                  & $\varphi_{asympt}$ & 0.0741 & 0.2541 & 0.6447 & 1.4299 & 2.2155 & 2.6103 & 2.8230 \\ \hline
{0.999}{}{} & $\varphi_{real}$ & 0.1518 & 0.34801 & 0.7407 &
1.5261 & 2.3115 & 2.7042 & 2.90065 \\ \cline{2-9}
                  & $\varphi_{asympt}$ & 0.1518 & 0.3480 & 0.7407 & 1.5261 & 2.3115 & 2.7042 & 2.9007 \\ \hline
\end{tabular}
\end{table}
\end{small}

\begin{small}
\begin{spacing}{0.05}
\begin{flushright}
%\caption{Table 4}\\
\textit{Table 4}
\end{flushright}
\begin{center}
%\caption{Table 2}\\
{$\varphi_{4}(\beta, \gamma)$}
\end{center}
\end{spacing}
\begin{table}[h!]
\begin{tabular}{|l|l|l|l|l|l|l|l|l|}
\hline
\begin{tabular}{ll}
 & $\gamma$\\
 $\beta$ &
\end{tabular} &  & $\pi/16$ & $\pi/8$ & $\pi/4$ & $\pi/2$ & $3\pi/4$ & $7\pi/8$ & $15\pi/16$ \\ \hline
{0.7}{}{} & $\varphi_{real}$ & undef. & undef. & undef. & 0.5437 &
1.379 & 1.6677 & 1.7554 \\ \cline{2-9}
                  & $\varphi_{asympt}$ & 12.9026 & -1.9623 & -0.1888 & 0.7647 & 1.3820 & 0.3939 & -10.153 \\ \hline
{0.8}{}{} & $\varphi_{real}$ & undef. & undef. & undef. & 0.8513 &
1.6312 & 1.9492 & 2.1627 \\ \cline{2-9}
                  & $\varphi_{asympt}$ & -6.5746 & -1.0249 & 0.0558 & 0.9250 & 1.6266 & 1.3313 & -3.8257 \\ \hline
{0.9}{}{} &$\varphi_{real}$  & undef. & undef. & 0.2315 & 1.1040 &
1.8824 & 2.2395 & 2.3764 \\ \cline{2-9}
                  & $\varphi_{asympt}$ & -1.9585 & -0.2709 & 0.3134 & 1.1221 & 1.8842 & 2.0853 & 0.7904 \\ \hline
{0.95}{}{} & $\varphi_{real}$ & undef. & undef. & 0.4512 & 1.2495
& 2.0322 & 2.41036 & 2.5730 \\ \cline{2-9}
                  & $\varphi_{asympt}$ & -0.5685 & 0.0205 & 0.4636 & 1.2552 & 2.0344 & 2.3767 & 2.1803 \\ \hline
{0.99}{}{} & $\varphi_{real}$ & undef. & 0.2470 & 0.6434 & 1.4294
& 2.2143 & 2.6059 & 2.7972 \\ \cline{2-9}
                  & $\varphi_{asympt}$ & 0.0364 & 0.2491 & 0.6439 & 1.4296 & 2.2147 & 2.6053 & 2.7853 \\ \hline
{0.999}{}{} & $\varphi_{real}$ & undef. & 0.3480 & 0.7407 & 1.5261
& 2.3115 & 2.7042 & 2.9005 \\ \cline{2-9}
                  & $\varphi_{asympt}$ & 0.1514 & 0.3479 & 0.7407 & 1.5261 & 2.3115 & 2.7042 & 2.9003 \\ \hline
\end{tabular}
\end{table}
\end{small}

\newpage
\begin{small}
\begin{spacing}{0.05}
\begin{flushright}
%\caption{Table 5}\\
\textit{Table 5}
\end{flushright}
\end{spacing}

\begin{table}[h!]
\begin{tabular}{|l|l|l|l|l|l|l|l|l|}
\hline
\begin{tabular}{ll}
 & $\gamma$\\
 $\beta$ &
\end{tabular} &  & $\pi/16$ & $\pi/8$ & $\pi/4$ & $\pi/2$ & $3\pi/4$ & $7\pi/8$ & $15\pi/16$ \\ \hline
{0.7}{}{} & $\delta_{\varphi_1}$ & 859\%  & 86\%  & 0.17\%  &
8.5\%  & - & - &-  \\ \cline{2-9}
                  & $\delta_{\varphi_2}$& 1423\%  & 151\%  & 15.6\%  & 3\%  & -  & - & - \\ \cline{2-9}
                  & $\delta_{\varphi_3}$ & - & -  & undef. & 7.6\%  & 13.7\%  & 75\%  & 543\%  \\ \cline{2-9}
                  & $\delta_{\varphi_4}$ & - & - & - & 41\%  & 0.21\%   & 76\% & 678\% \\ \hline
{0.8}{}{} & $\delta_{\varphi_1}$  & 541\% & 51\% & 0.33\% & 3.2\%
& - & - & - \\ \cline{2-9}
                  & $\delta_{\varphi_2}$ & 828\% & 85\% & 8.26\% & 1.7\% & 2.3\% & - & - \\ \cline{2-9}
                  & $\delta_{\varphi_3}$  & - & - & 27\% & 3.85\% & 6.4\% & 35\% & 272\% \\ \cline{2-9}
                  & $\delta_{\varphi_4}$  & - & - & - & 8.65\% & 0.28\% & 32\% & 277\% \\ \hline
{0.9}{}{} & $\delta_{\varphi_1}$ & 207\% & 17\% & 0.16\% & 0.89\%
& 2.8\% & - & - \\ \cline{2-9}
                  & $\delta_{\varphi_2}$ & 287\% & 24.9\% & 2.6\% & 0.7\% & 0.2\% & - & - \\ \cline{2-9}
                  & $\delta_{\varphi_3}$ & - & - & 1.5\% & 1.2\% & 1.6\% & 8.5\% & 65\% \\ \cline{2-9}
                  & $\delta_{\varphi_4}$  & - & - & 35\% & 1.6\% & 0.1\% & 6.9\% & 67\%  \\ \hline
{0.95}{}{} & $\delta_{\varphi_1}$ & 69\% & 4.6\% & 0.16\% & 0.3\%
& 0.4\% & - & - \\ \cline{2-9}
                  & $\delta_{\varphi_2}$ & 88\% & 6.8\% & 0.8\%  & 0.27\% & 0.06\%  & 0.64\% & - \\ \cline{2-9}
                  & $\delta_{\varphi_3}$ & - & 16\% & 0.34\% & 0.4\% & 0.4\% & 1.9\% & 16\% \\ \cline{2-9}
                  & $\delta_{\varphi_4}$  & - & - & 2.76\% & 0.46\% & 0.11\% & 1.4\% & 15\% \\ \hline
{0.99}{}{} & $\delta_{\varphi_1}$ & 3.4\% & 0.1\% & 0.05\% &
0.01\% & 0.02\% & - & - \\ \cline{2-9}
                  & $\delta_{\varphi_2}$ & 4\% & 0.29\% & 0.06\% & 0.045\% & 0.017\% & 0.2\% & -  \\ \cline{2-9}
                  & $\delta_{\varphi_3}$ & undef. & 0.2\% & 0.06\% & 0.03\% & 0.02\% & 0.06\%  & 0.48\% \\ \cline{2-9}
                  & $\delta_{\varphi_4}$  & - & 0.86\% & 0.08\% & 0.01\% & 0.019\% & 0.02\% & 0.4\% \\ \hline
{0.999}{}{} & $\delta_{\varphi_1}$ & 0.02\% & 0.015\% & 0.002\% &
0.001\% & 0.004\% & 0.003\% & - \\ \cline{2-9}
                  & $\delta_{\varphi_2}$ & 0.14\% & 0.026\% & 0.003\% & 0.0003 & 0.00003\% & 0.004\% & - \\ \cline{2-9}
                  & $\delta_{\varphi_3}$ & 0.02\%  & 0.001\% & 0.001\% & 0.0008\% & 0.0005\% & 0.0003\% & 0.003\% \\ \cline{2-9}
                  & $\delta_{\varphi_4}$ & - & 0.03\% & 0.002\% & 0.001\% & 0.0009\% & 0.002\% & 0.005\% \\ \hline

\end{tabular}
\end{table}
\end{small}

\end{document}